\documentclass[sigconf]{acmart}
\AtBeginDocument{%
  }

\setcopyright{acmlicensed}
\copyrightyear{2018}
\acmYear{2018}
\acmDOI{XXXXXXX.XXXXXXX}
\acmConference[Conference acronym 'XX]{Make sure to enter the correct
  conference title from your rights confirmation email}{June 03--05,
  2018}{Woodstock, NY}
\acmISBN{978-1-4503-XXXX-X/2018/06}

\begin{document}

\title{CHOIR: Chat-based Helper for Organizational Intelligence Repository}

\author{Sangwook Lee}
\email{sangwooklee@vt.edu}
\orcid{0000-0002-2600-4769}
\affiliation{%
  \institution{Virginia Tech}
  \city{Blacksburg}
  \state{Virginia}
  \country{USA}
}

\author{Adnan Abbas}
\email{adnana99@vt.edu}
\affiliation{%
  \institution{Virginia Tech}
  \city{Blacksburg}
  \state{Virginia}
  \country{USA}
}

\author{Yan Chen}
\email{ych@vt.edu}
\affiliation{%
  \institution{Virginia Tech}
  \city{Blacksburg}
  \state{Virginia}
  \country{USA}
}

\author{Sang Won Lee}
\email{sangwonlee@vt.edu}
\affiliation{%
  \institution{Virginia Tech}
  \city{Blacksburg}
  \state{Virginia}
  \country{USA}
}

\newcommand{\system}{CHOIR}

\begin{abstract}
Modern organizations frequently rely on chat-based platforms (e.g., Slack, Microsoft Teams, and Discord) for day-to-day communication and decision-making. As conversations evolve, organizational knowledge can get buried, prompting repeated searches and discussions. While maintaining shared documents, such as Wiki articles for the organization, offers a partial solution, it requires manual and timely efforts to keep it up to date, and it may not effectively preserve the social and contextual aspect of prior discussions. Moreover, reaching a consensus on document updates with relevant stakeholders can be time-consuming and complex. To address these challenges, we introduce CHOIR (Chat-based Helper for Organizational Intelligence Repository), a chatbot that integrates seamlessly with chat platforms. CHOIR automatically identifies and proposes edits to related documents, initiates discussions with relevant team members, and preserves contextual revision histories. By embedding knowledge management directly into chat environments and leveraging LLMs, CHOIR simplifies manual updates and supports consensus-driven editing based on maintained context with revision histories. We plan to design, deploy, and evaluate CHOIR in the context of maintaining an organizational memory for a research lab. We describe the chatbot's motivation, design, and early implementation to show how CHOIR streamlines collaborative document management.
\end{abstract}

\begin{CCSXML}
<ccs2012>
   <concept>
       <concept_id>10003120.10003130.10003233</concept_id>
       <concept_desc>Human-centered computing~Collaborative and social computing systems and tools</concept_desc>
       <concept_significance>500</concept_significance>
       </concept>
 </ccs2012>
\end{CCSXML}

\ccsdesc[500]{Human-centered computing~Collaborative and social computing systems and tools}

\keywords{chatbot, group chat, organizational memory, LLM, living document}


\begin{teaserfigure}
  \centering
  \includegraphics[width=\textwidth]{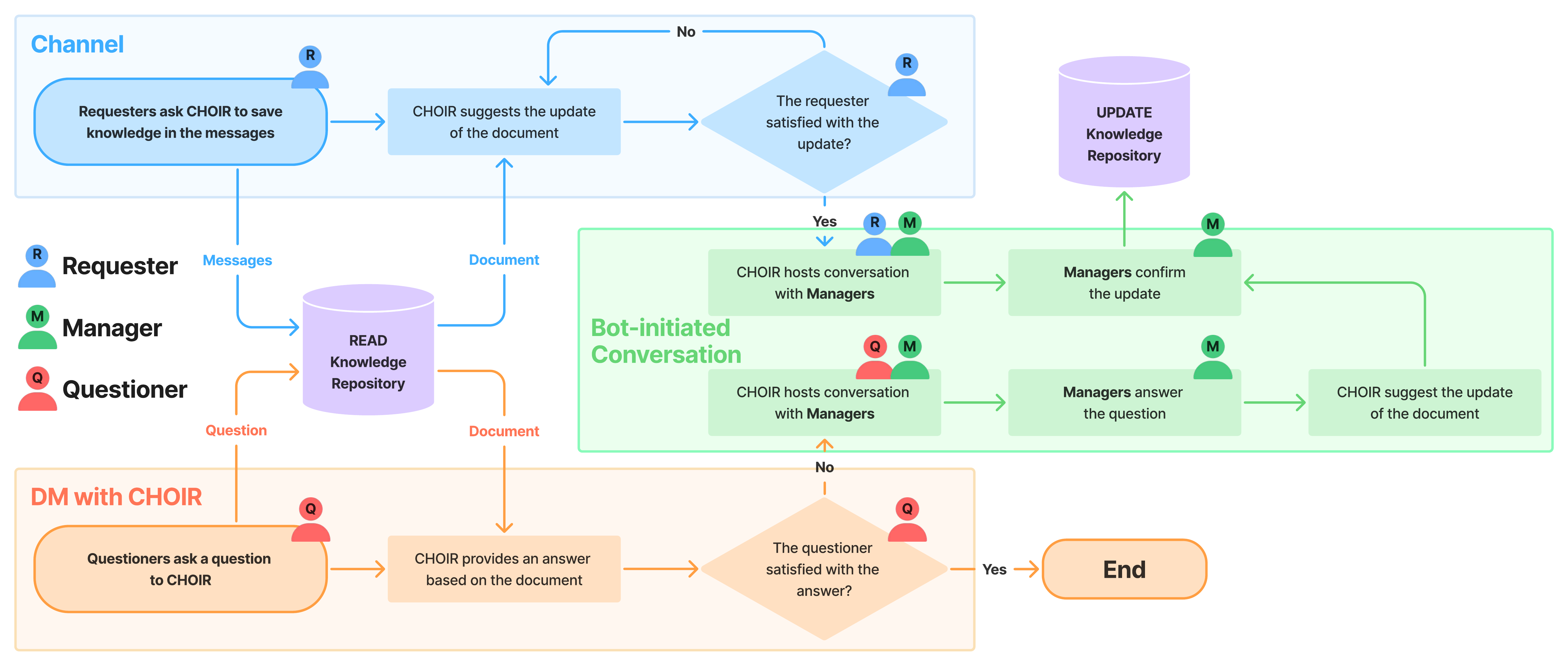}
  \caption{Workflow diagram of CHOIR}
  \Description{This flowchart shows how three user roles—Questioners, Requesters, and Managers—interact with CHOIR and the knowledge repository. The flow begins with a Requester, who interacts with CHOIR in a channel to preserve knowledge from ongoing conversations. The Requester asks CHOIR to store valuable information and may be prompted to confirm a proposed document update that reflects the newly gathered insights. If the Requester is satisfied with this update, it moves forward for Manager confirmation and is ultimately applied to the knowledge repository. Otherwise, the update process loops back into additional discussions. Meanwhile, a Questioner might directly message CHOIR with a query, prompting CHOIR to consult existing documents and provide an answer. If the Questioner is satisfied, the process ends. If not, CHOIR proposes a document update, inviting Managers to a initiated conversation. Managers then discuss and confirm any changes, which are also stored in the repository for future reference. Through these roles, CHOIR systematically captures and updates organizational knowledge in response to evolving needs.}
  \label{fig:system_flowchart}
\end{teaserfigure}

\maketitle

% \sang{I like what you did here to set the context of the platform that we are developing. Also, I wish there was one more paragraph to contextualize the work from an organizational memory perspective. Overall, when you contextualize the work, it should use literature. Having no citations for the first few paragraphs in the introduction is not a good sign. Also, I wish there was one concrete example that readers could relate to.}

\section{Introduction}
In modern organizations, group chat services such as Slack, Microsoft Teams, and Discord foster communication among members and facilitate collaboration and decision-making~\cite{handel2002chat, zhang2018making}. These platforms support channel-based communication and direct messaging, enabling smooth information exchange among teams or individuals when solving problems~\cite{wang2022group}. However, shared information or knowledge, which can benefit the organization if well maintained, can get easily buried in new messages~\cite{zhang2018making} in a dynamic and fast-paced chat environment~\cite{cameron2005unintended}. As a result, users often fail to utilize existing knowledge in similar problem-solving situations or end up searching through past conversations and repeating discussions that have already taken place~\cite{zhang2018making}.

Researchers in CSCW have addressed the challenge of maintaining knowledge that emerges from an organization, which is described as organizational memory~\cite{ackerman2004organizational, ackerman2013sharing}. Organizational memory refers to the knowledge stored in an organization regarding decision-making stimulus and outcomes, which can be reused to inform future decisions~\cite{walsh2009organizational}. The line of research has examined how repository systems can help members of an organization to store and retrieve knowledge~\cite{ackerman1990answer, stein1995actualizing, burke1997question, pipek2003pruning, halverson2004behind, schaffert2006ikewiki, grudin2010wikis}. AnswerGarden was one of the early systems that allowed users to explore the tree of organizational memory to find the information they needed; if the information was missing, they could email relevant experts to expand the tree~\cite{ackerman1990answer}. Other scholars have also studied wiki systems as another means to store and share knowledge within companies~\cite{schaffert2006ikewiki,grudin2010wikis}. Recently, with the rising popularity of cloud-based repository services like Google Drive and Dropbox, organizations can more easily maintain and share knowledge base online~\cite{saratchandra2022role}. Despite this convenience, the omission of social relationships, and the risk of losing rich context once it is converted into documents remains fundamental limitations for knowledge repositories~\cite{ackerman2013sharing}.

Recent advances in LLMs have the potential to overcome the limitations of repository-based models. They can summarize key insights from lengthy texts, such as news articles~\cite{zhang2024benchmarking}, group dialogues~\cite{asthana2023summaries, tian2024dialogue}, and then integrate those insights into existing material through chat interactions~\cite{laban2024beyond}. They can also serve as moderators~\cite{bhatia2024moderating} or debaters~\cite{chiang2024enhancing} that interpret and guide the flow of dialogue among individuals. RAG-based solutions rely on an LLM's ability to provide context-aware answers rooted in existing documents~\cite{lewis2020retrieval}, and various organizations have already adopted them~\cite{khangarotCouncilPostRAG}. Leveraging LLM's advanced capabilities, we designed \textbf{CHOIR} (Chat-based Helper for Organizational Intelligence Repository) to seamlessly capture knowledge within group chats, foster collaborative organizational memory, and provide contextually relevant responses to users. We have three design goals as follows: 

\begin{itemize}
    \item {Facilitate the document update process by bridging  conversations and persistent documentation}
    \item {Encourage discussion within a group-chat environment among Stakeholders for Document Updates}
    \item {Help users grasp the document's evolving context by highlighting version history and conversation logs}
\end{itemize}

We are currently developing CHOIR as a Slack application and aim to explore the following research questions:

\begin{itemize}
    \item {\textbf{RQ1}: How does chatbot-driven document update automation reduce management effort (e.g., time or cognitive load) while fostering user contributions in collaborative document tasks?}
    \item {\textbf{RQ2}: How does initiating discussion around documents via a chatbot influence how team members discuss and negotiate document changes, and what challenges and opportunities emerge for integrating those updates?}
    % \item {\textbf{RQ3}: How does contextual information stored in revision histories (e.g., conversation logs or metadata) affect chatbot-assisted Q\&A and collaborative discussions, ultimately shaping the final document outcome?}
\end{itemize}

\section{System Design}
\begin{figure*}
  \centering
  \includegraphics[width=\textwidth]{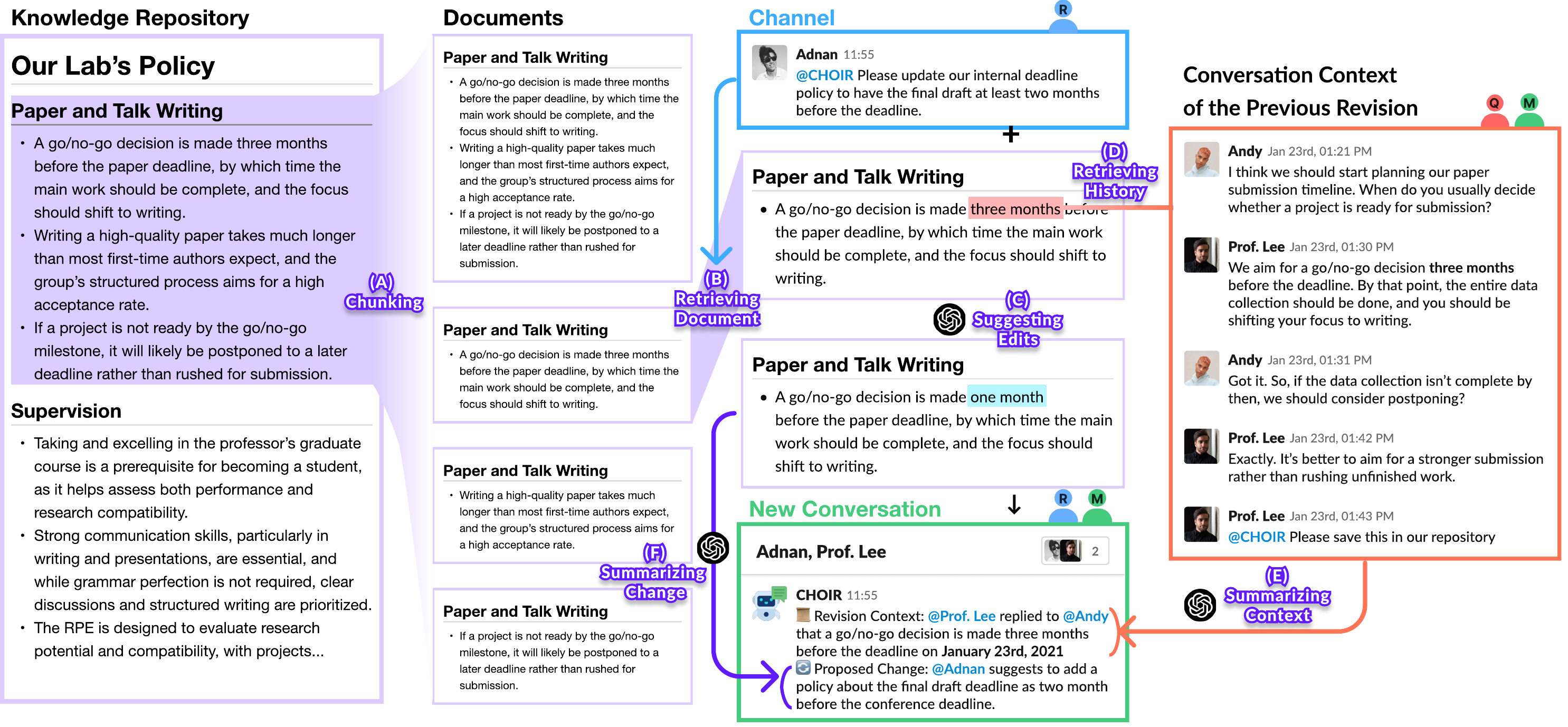}
  \caption{An overview of CHOIR's pipeline for chatbot-driven document updates and contextual revision management. (A) CHOIR segments repository knowledge into smaller chunks for efficient retrieval. (B) It retrieves the relevant document based on the Requester's channel mention. (C) A suggested edit is provided, derived from the Requester's selected message. (D) Once the Requester decides to apply the edit, prior revision history (e.g., a conversation between Questioner and Manager) is retrieved to offer contextual background. (E) CHOIR summarizes this context during the conversation it initiates. (F) It also summarizes the proposed change for Managers to review and confirm, ensuring they understand its background before approving.}
  \Description{This figure illustrates CHOIR's pipeline for managing document updates and revision context in a chat environment. On the left, ``Knowledge Repository'' contains various policy documents. In the middle, labeled ``Documents,'' the system (A) chunks these documents into smaller parts for efficient retrieval, and (B) locates the relevant document when a Requester mentions CHOIR in the channel. Next, (C) a suggested edit is proposed based on the Requester's selected message. If the user wants to proceed, (D) prior revision history is retrieved to provide context. CHOIR then (E) summarizes this context in the conversation it initiates, and (F) summarizes the change for Managers so they can understand the background and confirm the update. On the right, the ``Conversation Context of the Previous Revision'' box shows an example of how user discussions are logged, demonstrating how CHOIR bridges the repository and the chat to refine and store knowledge.}
  \label{fig:system_pipeline}
\end{figure*}

\subsection{Design Goals}
Prior research highlights the challenges of traditional repository models such as the difficulty of continuous maintenance, the limited reflection of social relationships, and the potential loss of rich context remain~\cite{ackerman2013sharing}. We show how our system's design seeks to overcome these issues by 1) bridging conversations with persistent documentation, 2) encourging discussion among stakeholders, and 3) preserving document's context with revision histories.

\subsubsection{Facilitate the document update process by bridging conversations and persistent documentation}

Our system is designed to capture knowledge and information on demand, leveraging conversations into persistent documents, thereby ensuring that organizational memory is easily and reliably maintained. A study on early groupware adoption found that organizations struggled to motivate members to contribute to share information, highlighting the importance of reward systems for contributions~\cite{orlikowski1992learning}. One reason is that continuously updating organizational memory requires manual efforts to keep it up-to-date. 

Research collectively has highlighted the importance of bridging conversations and persistent documentation to better facilitate the continuous update and management of organizational knowledge. Conversations among members can be excellent material for building organizational memory. One early system, Answer Garden, actively used email to broaden its stored knowledge~\cite{ackerman1990answer}. Even in Wikipedia—an online form of organizational memory—contributors share opinions about edits via article talk pages~\cite{kittur2008harnessing, schneider2011understanding}. There has also been extensive research on systems that capture and organize important conversation details~\cite{nam2007arkose, zhang2017wikum, zhang2018making}. To that end, we will facilitate members to update their organizational memory using conversations.

% \subsubsection{Encourage instant and context-sensitive collaboration by initiating instant discussion around documents} % context-preserving when updating and providing the communication context when facilitating discussion. 

\subsubsection{Encourage Discussion within a group-chat environment among Stakeholders for Document Updates}

CHOIR is designed to foster instant and context-sensitive collaboration by providing a communication channel where users can instantly discuss information tied directly to the document content. Only using a repository for organizational memory often fails to capture social contexts, such as the rationale behind the changes, the author's expertise, the reliability of the information, and whether it is officially confirmed~\cite{ackerman2013sharing}. In contrast, a group chat environment, such as Slack, contains rich information about the suggested changes and allows members to share and verify it immediately. For example, users can directly ask relevant experts about a document's information and better understand its context for specific situations. Such a conversation also helps mitigate conflicts among contributors and fosters a more cooperative culture. On Wikipedia, for instance, new editors often see their contributions reverted without clear reasons, which discourages them~\cite{halfaker2011don}, but help from experienced editors significantly improves newcomer retention~\cite{morgan2013tea}. Based on these previous findings, we designed our system to prompt conversation around documents so that users can account for the social context and receive assistance with document updates.

\subsubsection{Help users grasp the document's evolving context by highlighting version histories}

We design the system to preserve and showcase both version history and conversation logs, enabling users to more easily understand the evolving context behind each document update. When adding new information to organizational memory, authors often omit details that can contextualize the changes~\cite{ackerman1999organizational}. Such context may include social context, the source of the information, or the circumstances under which decisions were made. While decontextualizing the information can make it compact and thus easy to reuse in the future, retaining certain contextual details can be advantageous when they become crucial for specific decision-making scenarios~\cite{young2006don}. In this regard, preserving conversation logs and the version history from the document updating process can be highly effective for capturing the document's evolving context. Consequently, we designed our system to store these revision histories as contextual information when updating documents. 
We will present this information to ensure that users can readily understand the context behind the knowledge contained in the document. 

\subsection{System Features}

\subsubsection{Chatbot-driven document update}
\begin{figure}
  \centering
  \includegraphics[width=\linewidth]{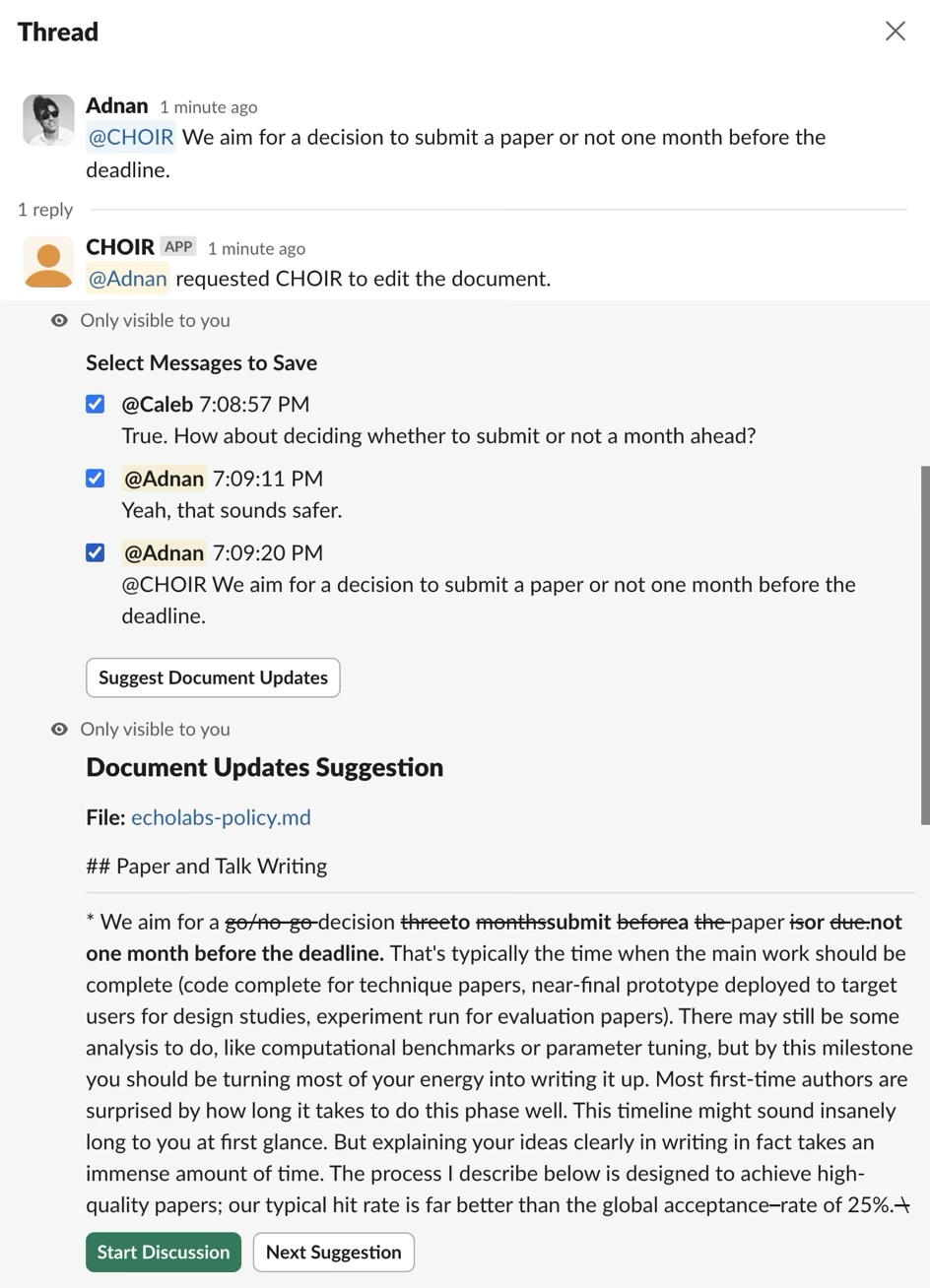}
  
  \caption{User Interface for Chatbot-driven document update}
  \Description{User Interface for Chatbot-driven document update
Adnan: @CHOIR We aim for a decision to submit a paper or not one month before the deadline.

CHOIR: @Adnan requested CHOIR to edit the document.
Only visible to you

Select Messages to Save

@Caleb: True. How about deciding whether to submit or not a month ahead?

@Adnan: Yeah, that sounds safer.

@Adnan: @CHOIR We aim for a decision to submit a paper or not one month before the deadline.

Only visible to you

Document Updates Suggestion
File: echolabs-policy.md
## Paper and Talk Writing
* We aim for a go/no-go decision threeto monthssubmit beforea the paper isor due.not one month before the deadline. That's typically the time when the main work should be complete (code complete for technique papers, near-final prototype deployed to target users for design studies, experiment run for evaluation papers). There may still be some analysis to do, like computational benchmarks or parameter tuning, but by this milestone you should be turning most of your energy into writing it up. Most first-time authors are surprised by how long it takes to do this phase well. This timeline might sound insanely long to you at first glance. But explaining your ideas clearly in writing in fact takes an immense amount of time. The process I describe below is designed to achieve high-quality papers; our typical hit rate is far better than the global acceptance  rate of 25

Two buttons with Start Discussion and Next Suggestion
  }
  \label{fig:update_document}
\end{figure}

In Figure~\ref{fig:system_flowchart}, when a \textit{Requester} wishes to store knowledge from a channel message, they can invoke CHOIR by mentioning it in either a new message or a thread reply. CHOIR then interacts with the user to suggest how to update the relevant document in the repository. The workflow is as follows: 1) The user selects one of the previously shown messages that contain the information they want to save. 2) As shown in Figure~\ref{fig:system_pipeline}, CHOIR retrieves the most relevant document from the repository. 3) CHOIR uses LLM prompting to edit the retrieved document based on the selected message. If the \textit{Requester} is not satisfied with the proposed update, they can press a button to see the next suggestion or create a new document altogether (Figure~\ref{fig:update_document}). 

\subsubsection{Initiating discussion around documents}
In Figure~\ref{fig:system_flowchart}, CHOIR can also host a conversation by instantly bringing the \textit{Requester} and a \textit{Manager} (the person with permission to update the repository) together. When starting this conversation, CHOIR explains what update has been proposed and provides context from the revision message, as seen in Figure~\ref{fig:system_pipeline}, ensuring both the \textit{Requester} and \textit{Manager} have sufficient information to begin discussing the update. After the discussion, the \textit{Manager} can approve or reject the proposed change.

CHOIR can similarly host instant conversations when a \textit{Questioner} asks it a question. A user can direct-message CHOIR with a query, and CHOIR will respond based on documents in the repository. If the user finds the answer insufficient, CHOIR can invite both the \textit{Questioner} and a \textit{Manager} to a conversation. In this chat, the \textit{Manager} can clarify the answer, and the user can choose to update the document through CHOIR if necessary.

\subsubsection{Contextual revision history}
When CHOIR updates a document, it saves the user-selected messages as as part of the revision history, thereby capturing important context. These messages can include information about events at the time of the update, the source of the content, or the person involved in the conversation. CHOIR also uses LLM prompting to summarize this contextual information and present it during the hosted conversation, making it readily available for all participants. For example, in Figure ~\ref{fig:system_pipeline}, Andy is not a \textit{Manager} and was not initially invited to the conversation, but Prof. Lee can invite Andy to gather his opinion on Adnan's proposed update. This allows for more comprehensive collaboration and discussion around document changes.

\subsection{System Implementation}
The CHOIR server is built using Slack Bolt~\cite{bolt} and Block Kit~\cite{blockkit} frameworks, and it uses the GitHub REST API~\cite{githubapi} to read and update files in a GitHub repository. The knowledge in the GitHub repository is stored as markdown files, which members can directly view and edit through an integrated Gitbook~\cite{gitbook}. CHOIR parses the markdown files and stores them as LangChain~\cite{langchain} Document objects every time the repository is updated. These stored objects are then used for document updates or for searching relevant documents to answer user queries, based on vector embeddings. When updating a document, CHOIR includes the messages used in the document as metadata in the commit message and pushes it to the GitHub repository. Later, when CHOIR needs to retrieve a file to update, it uses the \texttt{git history} command to fetch the commit messages. The messages in these commit messages are used to provide context for updates in the instant conversation.

\section{Future Plan}

In our future work, we will conduct a formative study to uncover current practices and challenges in organizational memory management and revisit our design goals by incorporating real-world practices. After finishing the system development, we will deploy our system for a field study to evaluate CHOIR in a real-world setting.

\subsection{Formative Study}

We plan to begin with a formative study that will help us gain a detailed understanding of how members manage their organizational memory and the challenges they face. We have chosen a university research lab as our target because labs often experience frequent member turnover, depend on regular communication, and are more accessible for interviews and observations than corporations. In these labs, we will interview students or supervisors to learn about the types of documents they create, the individuals or roles responsible for maintaining them, and how they update and use these documents in their daily workflows. We will also explore any difficulties they encounter during the document management process. By analyzing the interview data, we aim to refine CHOIR's design goals to align more closely with the real-world needs of our participants and add features that directly address the challenges they report.

\subsection{System Evaluation: Field Study}
After refining the system, we will conduct a field study to evaluate CHOIR in the actual research lab environment and address our research questions. During the study, lab members will deploy CHOIR in their Slack workspaces for document management and updates. We will monitor the usage logs to see how participants engage with CHOIR, tracking aspects like the frequency of chatbot interactions, the types of tasks performed, and how effectively it supports collaboration for maintaining organizational memory. Additionally, we will investigate their experience compared to their existing practice of organizational memory, if any.
% these findings to baseline data from periods when the labs managed their documents without CHOIR. 
Through this comparison, we can assess the impact of CHOIR on existing workflows, document quality, and overall collaborative practices. 

\bibliographystyle{ACM-Reference-Format}
\bibliography{references}


\begin{thebibliography}{37}

%%% the Web address.

\ifx \showCODEN    \undefined \def \showCODEN     #1{\unskip}     \fi
\ifx \showISBNx    \undefined \def \showISBNx     #1{\unskip}     \fi
\ifx \showISBNxiii \undefined \def \showISBNxiii  #1{\unskip}     \fi
\ifx \showISSN     \undefined \def \showISSN      #1{\unskip}     \fi
\ifx \showLCCN     \undefined \def \showLCCN      #1{\unskip}     \fi
\ifx \shownote     \undefined \def \shownote      #1{#1}          \fi
\ifx \showarticletitle \undefined \def \showarticletitle #1{#1}   \fi
\ifx \showURL      \undefined \def \showURL       {\relax}        \fi
% invisible to TeX
\providecommand\bibfield[2]{#2}
\providecommand\bibinfo[2]{#2}
\providecommand\natexlab[1]{#1}
\providecommand\showeprint[2][]{arXiv:#2}

\bibitem[Ackerman et~al\mbox{.}(2013)]%
        {ackerman2013sharing}
\bibfield{author}{\bibinfo{person}{Mark~S Ackerman}, \bibinfo{person}{Juri Dachtera}, \bibinfo{person}{Volkmar Pipek}, {and} \bibinfo{person}{Volker Wulf}.} \bibinfo{year}{2013}\natexlab{}.
\newblock \showarticletitle{Sharing knowledge and expertise: The CSCW view of knowledge management}.
\newblock \bibinfo{journal}{\emph{Computer Supported Cooperative Work (CSCW)}}  \bibinfo{volume}{22} (\bibinfo{year}{2013}), \bibinfo{pages}{531--573}.
\newblock


\bibitem[Ackerman and Halverson(1999)]%
        {ackerman1999organizational}
\bibfield{author}{\bibinfo{person}{Mark~S Ackerman} {and} \bibinfo{person}{Christine Halverson}.} \bibinfo{year}{1999}\natexlab{}.
\newblock \showarticletitle{Organizational memory: processes, boundary objects, and trajectories}. In \bibinfo{booktitle}{\emph{Proceedings of the 32nd Annual Hawaii International Conference on Systems Sciences. 1999. HICSS-32. Abstracts and CD-ROM of Full Papers}}. IEEE, \bibinfo{pages}{12--pp}.
\newblock


\bibitem[Ackerman and Halverson(2004)]%
        {ackerman2004organizational}
\bibfield{author}{\bibinfo{person}{Mark~S Ackerman} {and} \bibinfo{person}{Christine Halverson}.} \bibinfo{year}{2004}\natexlab{}.
\newblock \showarticletitle{Organizational memory as objects, processes, and trajectories: An examination of organizational memory in use}.
\newblock \bibinfo{journal}{\emph{Computer Supported Cooperative Work (CSCW)}}  \bibinfo{volume}{13} (\bibinfo{year}{2004}), \bibinfo{pages}{155--189}.
\newblock


\bibitem[Ackerman and Malone(1990)]%
        {ackerman1990answer}
\bibfield{author}{\bibinfo{person}{Mark~S Ackerman} {and} \bibinfo{person}{Thomas~W Malone}.} \bibinfo{year}{1990}\natexlab{}.
\newblock \showarticletitle{Answer Garden: A tool for growing organizational memory}.
\newblock \bibinfo{journal}{\emph{ACM SIGOIS Bulletin}} \bibinfo{volume}{11}, \bibinfo{number}{2-3} (\bibinfo{year}{1990}), \bibinfo{pages}{31--39}.
\newblock


\bibitem[Asthana et~al\mbox{.}(2023)]%
        {asthana2023summaries}
\bibfield{author}{\bibinfo{person}{Sumit Asthana}, \bibinfo{person}{Sagih Hilleli}, \bibinfo{person}{Pengcheng He}, {and} \bibinfo{person}{Aaron Halfaker}.} \bibinfo{year}{2023}\natexlab{}.
\newblock \showarticletitle{Summaries, Highlights, and Action items: Design, implementation and evaluation of an LLM-powered meeting recap system}.
\newblock \bibinfo{journal}{\emph{arXiv preprint arXiv:2307.15793}} (\bibinfo{year}{2023}).
\newblock


\bibitem[Bhatia and Sukthankar(2024)]%
        {bhatia2024moderating}
\bibfield{author}{\bibinfo{person}{Aaditya Bhatia} {and} \bibinfo{person}{Gita Sukthankar}.} \bibinfo{year}{2024}\natexlab{}.
\newblock \showarticletitle{Moderating Democratic Discourse with LLMs}. In \bibinfo{booktitle}{\emph{International Conference on Social Computing, Behavioral-Cultural Modeling and Prediction and Behavior Representation in Modeling and Simulation}}. Springer, \bibinfo{pages}{123--132}.
\newblock


\bibitem[Burke et~al\mbox{.}(1997)]%
        {burke1997question}
\bibfield{author}{\bibinfo{person}{Robin~D Burke}, \bibinfo{person}{Kristian~J Hammond}, \bibinfo{person}{Vladimir Kulyukin}, \bibinfo{person}{Steven~L Lytinen}, \bibinfo{person}{Noriko Tomuro}, {and} \bibinfo{person}{Scott Schoenberg}.} \bibinfo{year}{1997}\natexlab{}.
\newblock \showarticletitle{Question answering from frequently asked question files: Experiences with the faq finder system}.
\newblock \bibinfo{journal}{\emph{AI magazine}} \bibinfo{volume}{18}, \bibinfo{number}{2} (\bibinfo{year}{1997}), \bibinfo{pages}{57--57}.
\newblock


\bibitem[Cameron and Webster(2005)]%
        {cameron2005unintended}
\bibfield{author}{\bibinfo{person}{Ann~Frances Cameron} {and} \bibinfo{person}{Jane Webster}.} \bibinfo{year}{2005}\natexlab{}.
\newblock \showarticletitle{Unintended consequences of emerging communication technologies: Instant messaging in the workplace}.
\newblock \bibinfo{journal}{\emph{Computers in Human behavior}} \bibinfo{volume}{21}, \bibinfo{number}{1} (\bibinfo{year}{2005}), \bibinfo{pages}{85--103}.
\newblock


\bibitem[Chiang et~al\mbox{.}(2024)]%
        {chiang2024enhancing}
\bibfield{author}{\bibinfo{person}{Chun-Wei Chiang}, \bibinfo{person}{Zhuoran Lu}, \bibinfo{person}{Zhuoyan Li}, {and} \bibinfo{person}{Ming Yin}.} \bibinfo{year}{2024}\natexlab{}.
\newblock \showarticletitle{Enhancing AI-Assisted Group Decision Making through LLM-Powered Devil's Advocate}. In \bibinfo{booktitle}{\emph{Proceedings of the 29th International Conference on Intelligent User Interfaces}}. \bibinfo{pages}{103--119}.
\newblock


\bibitem[Gitbook(2025)]%
        {gitbook}
\bibfield{author}{\bibinfo{person}{Inc. Gitbook}.} \bibinfo{year}{2025}\natexlab{}.
\newblock \bibinfo{booktitle}{\emph{Gitbook}}.
\newblock
\urldef\tempurl%
\url{https://www.gitbook.com/}
\showURL{%
\tempurl}


\bibitem[GitHub(2025)]%
        {githubapi}
\bibfield{author}{\bibinfo{person}{Inc. GitHub}.} \bibinfo{year}{2025}\natexlab{}.
\newblock \bibinfo{booktitle}{\emph{GitHub REST API}}.
\newblock
\urldef\tempurl%
\url{https://docs.github.com/en/rest}
\showURL{%
\tempurl}


\bibitem[Grudin and Poole(2010)]%
        {grudin2010wikis}
\bibfield{author}{\bibinfo{person}{Jonathan Grudin} {and} \bibinfo{person}{Erika~Shehan Poole}.} \bibinfo{year}{2010}\natexlab{}.
\newblock \showarticletitle{Wikis at work: success factors and challenges for sustainability of enterprise Wikis}. In \bibinfo{booktitle}{\emph{Proceedings of the 6th international symposium on Wikis and open collaboration}}. \bibinfo{pages}{1--8}.
\newblock


\bibitem[Halfaker et~al\mbox{.}(2011)]%
        {halfaker2011don}
\bibfield{author}{\bibinfo{person}{Aaron Halfaker}, \bibinfo{person}{Aniket Kittur}, {and} \bibinfo{person}{John Riedl}.} \bibinfo{year}{2011}\natexlab{}.
\newblock \showarticletitle{Don't bite the newbies: how reverts affect the quantity and quality of Wikipedia work}. In \bibinfo{booktitle}{\emph{Proceedings of the 7th international symposium on wikis and open collaboration}}. \bibinfo{pages}{163--172}.
\newblock


\bibitem[Halverson et~al\mbox{.}(2004)]%
        {halverson2004behind}
\bibfield{author}{\bibinfo{person}{Christine~A Halverson}, \bibinfo{person}{Thomas Erickson}, {and} \bibinfo{person}{Mark~S Ackerman}.} \bibinfo{year}{2004}\natexlab{}.
\newblock \showarticletitle{Behind the help desk: evolution of a knowledge management system in a large organization}. In \bibinfo{booktitle}{\emph{Proceedings of the 2004 ACM conference on Computer supported cooperative work}}. \bibinfo{pages}{304--313}.
\newblock


\bibitem[Handel and Herbsleb(2002)]%
        {handel2002chat}
\bibfield{author}{\bibinfo{person}{Mark Handel} {and} \bibinfo{person}{James~D Herbsleb}.} \bibinfo{year}{2002}\natexlab{}.
\newblock \showarticletitle{What is chat doing in the workplace?}. In \bibinfo{booktitle}{\emph{Proceedings of the 2002 ACM conference on Computer supported cooperative work}}. \bibinfo{pages}{1--10}.
\newblock


\bibitem[Khangarot({[n.\,d.]})]%
        {khangarotCouncilPostRAG}
\bibfield{author}{\bibinfo{person}{Samder Khangarot}.} \bibinfo{year}{[n.\,d.]}\natexlab{}.
\newblock \bibinfo{title}{Council {{Post}}: {{The RAG Effect}}: {{How AI Is Becoming More Relevant And Accurate}}}.
\newblock \bibinfo{howpublished}{https://www.forbes.com/councils/forbesbusinesscouncil/2024/04/24/the-rag-effect-how-ai-is-becoming-more-relevant-and-accurate/}.
\newblock


\bibitem[Kittur and Kraut(2008)]%
        {kittur2008harnessing}
\bibfield{author}{\bibinfo{person}{Aniket Kittur} {and} \bibinfo{person}{Robert~E Kraut}.} \bibinfo{year}{2008}\natexlab{}.
\newblock \showarticletitle{Harnessing the wisdom of crowds in wikipedia: quality through coordination}. In \bibinfo{booktitle}{\emph{Proceedings of the 2008 ACM conference on Computer supported cooperative work}}. \bibinfo{pages}{37--46}.
\newblock


\bibitem[Laban et~al\mbox{.}(2024)]%
        {laban2024beyond}
\bibfield{author}{\bibinfo{person}{Philippe Laban}, \bibinfo{person}{Jesse Vig}, \bibinfo{person}{Marti Hearst}, \bibinfo{person}{Caiming Xiong}, {and} \bibinfo{person}{Chien-Sheng Wu}.} \bibinfo{year}{2024}\natexlab{}.
\newblock \showarticletitle{Beyond the chat: Executable and verifiable text-editing with llms}. In \bibinfo{booktitle}{\emph{Proceedings of the 37th Annual ACM Symposium on User Interface Software and Technology}}. \bibinfo{pages}{1--23}.
\newblock


\bibitem[LangChain(2025)]%
        {langchain}
\bibfield{author}{\bibinfo{person}{LangChain}.} \bibinfo{year}{2025}\natexlab{}.
\newblock \bibinfo{booktitle}{\emph{LangChain}}.
\newblock
\urldef\tempurl%
\url{https://www.langchain.com/}
\showURL{%
\tempurl}


\bibitem[Lewis et~al\mbox{.}(2020)]%
        {lewis2020retrieval}
\bibfield{author}{\bibinfo{person}{Patrick Lewis}, \bibinfo{person}{Ethan Perez}, \bibinfo{person}{Aleksandra Piktus}, \bibinfo{person}{Fabio Petroni}, \bibinfo{person}{Vladimir Karpukhin}, \bibinfo{person}{Naman Goyal}, \bibinfo{person}{Heinrich K{\"u}ttler}, \bibinfo{person}{Mike Lewis}, \bibinfo{person}{Wen-tau Yih}, \bibinfo{person}{Tim Rockt{\"a}schel}, {et~al\mbox{.}}} \bibinfo{year}{2020}\natexlab{}.
\newblock \showarticletitle{Retrieval-augmented generation for knowledge-intensive nlp tasks}.
\newblock \bibinfo{journal}{\emph{Advances in Neural Information Processing Systems}}  \bibinfo{volume}{33} (\bibinfo{year}{2020}), \bibinfo{pages}{9459--9474}.
\newblock


\bibitem[Morgan et~al\mbox{.}(2013)]%
        {morgan2013tea}
\bibfield{author}{\bibinfo{person}{Jonathan~T Morgan}, \bibinfo{person}{Siko Bouterse}, \bibinfo{person}{Heather Walls}, {and} \bibinfo{person}{Sarah Stierch}.} \bibinfo{year}{2013}\natexlab{}.
\newblock \showarticletitle{Tea and sympathy: crafting positive new user experiences on wikipedia}. In \bibinfo{booktitle}{\emph{Proceedings of the 2013 conference on Computer supported cooperative work}}. \bibinfo{pages}{839--848}.
\newblock


\bibitem[Nam and Ackerman(2007)]%
        {nam2007arkose}
\bibfield{author}{\bibinfo{person}{Kevin~K Nam} {and} \bibinfo{person}{Mark~S Ackerman}.} \bibinfo{year}{2007}\natexlab{}.
\newblock \showarticletitle{Arkose: reusing informal information from online discussions}. In \bibinfo{booktitle}{\emph{Proceedings of the 2007 ACM International Conference on Supporting Group Work}}. \bibinfo{pages}{137--146}.
\newblock


\bibitem[Orlikowski(1992)]%
        {orlikowski1992learning}
\bibfield{author}{\bibinfo{person}{Wanda~J Orlikowski}.} \bibinfo{year}{1992}\natexlab{}.
\newblock \showarticletitle{Learning from notes: Organizational issues in groupware implementation}. In \bibinfo{booktitle}{\emph{Proceedings of the 1992 ACM conference on Computer-supported cooperative work}}. \bibinfo{pages}{362--369}.
\newblock


\bibitem[Pipek and Wulf(2003)]%
        {pipek2003pruning}
\bibfield{author}{\bibinfo{person}{Volkmar Pipek} {and} \bibinfo{person}{Volker Wulf}.} \bibinfo{year}{2003}\natexlab{}.
\newblock \showarticletitle{Pruning the answer garden: knowledge sharing in maintenance engineering}. In \bibinfo{booktitle}{\emph{ECSCW 2003: Proceedings of the Eighth European Conference on Computer Supported Cooperative Work 14--18 September 2003, Helsinki, Finland}}. Springer, \bibinfo{pages}{1--20}.
\newblock


\bibitem[Saratchandra and Shrestha(2022)]%
        {saratchandra2022role}
\bibfield{author}{\bibinfo{person}{Minu Saratchandra} {and} \bibinfo{person}{Anup Shrestha}.} \bibinfo{year}{2022}\natexlab{}.
\newblock \showarticletitle{The role of cloud computing in knowledge management for small and medium enterprises: a systematic literature review}.
\newblock \bibinfo{journal}{\emph{Journal of Knowledge Management}} \bibinfo{volume}{26}, \bibinfo{number}{10} (\bibinfo{year}{2022}), \bibinfo{pages}{2668--2698}.
\newblock


\bibitem[Schaffert(2006)]%
        {schaffert2006ikewiki}
\bibfield{author}{\bibinfo{person}{Sebastian Schaffert}.} \bibinfo{year}{2006}\natexlab{}.
\newblock \showarticletitle{IkeWiki: A semantic wiki for collaborative knowledge management}. In \bibinfo{booktitle}{\emph{15th IEEE International Workshops on Enabling Technologies: Infrastructure for Collaborative Enterprises (WETICE'06)}}. IEEE, \bibinfo{pages}{388--396}.
\newblock


\bibitem[Schneider et~al\mbox{.}(2011)]%
        {schneider2011understanding}
\bibfield{author}{\bibinfo{person}{Jodi Schneider}, \bibinfo{person}{Alexandre Passant}, {and} \bibinfo{person}{John~G Breslin}.} \bibinfo{year}{2011}\natexlab{}.
\newblock \showarticletitle{Understanding and improving Wikipedia article discussion spaces}. In \bibinfo{booktitle}{\emph{Proceedings of the 2011 ACM Symposium on Applied Computing}}. \bibinfo{pages}{808--813}.
\newblock


\bibitem[Stein and Zwass(1995)]%
        {stein1995actualizing}
\bibfield{author}{\bibinfo{person}{Eric~W Stein} {and} \bibinfo{person}{Vladimir Zwass}.} \bibinfo{year}{1995}\natexlab{}.
\newblock \showarticletitle{Actualizing organizational memory with information systems}.
\newblock \bibinfo{journal}{\emph{Information systems research}} \bibinfo{volume}{6}, \bibinfo{number}{2} (\bibinfo{year}{1995}), \bibinfo{pages}{85--117}.
\newblock


\bibitem[Technologies(2025a)]%
        {blockkit}
\bibfield{author}{\bibinfo{person}{Slack Technologies}.} \bibinfo{year}{2025}\natexlab{a}.
\newblock \bibinfo{booktitle}{\emph{Slack Block Kit}}.
\newblock
\urldef\tempurl%
\url{https://api.slack.com/block-kit}
\showURL{%
\tempurl}


\bibitem[Technologies(2025b)]%
        {bolt}
\bibfield{author}{\bibinfo{person}{Slack Technologies}.} \bibinfo{year}{2025}\natexlab{b}.
\newblock \bibinfo{booktitle}{\emph{Slack Bolt}}.
\newblock
\urldef\tempurl%
\url{https://api.slack.com/bolt}
\showURL{%
\tempurl}


\bibitem[Tian et~al\mbox{.}(2024)]%
        {tian2024dialogue}
\bibfield{author}{\bibinfo{person}{Yuanhe Tian}, \bibinfo{person}{Fei Xia}, {and} \bibinfo{person}{Yan Song}.} \bibinfo{year}{2024}\natexlab{}.
\newblock \showarticletitle{Dialogue summarization with mixture of experts based on large language models}. In \bibinfo{booktitle}{\emph{Proceedings of the 62nd Annual Meeting of the Association for Computational Linguistics (Volume 1: Long Papers)}}. \bibinfo{pages}{7143--7155}.
\newblock


\bibitem[Walsh and Ungson(2009)]%
        {walsh2009organizational}
\bibfield{author}{\bibinfo{person}{James~P Walsh} {and} \bibinfo{person}{Gerardo~Rivera Ungson}.} \bibinfo{year}{2009}\natexlab{}.
\newblock \showarticletitle{Organizational memory}.
\newblock In \bibinfo{booktitle}{\emph{Knowledge in Organisations}}. \bibinfo{publisher}{Routledge}, \bibinfo{pages}{177--212}.
\newblock


\bibitem[Wang et~al\mbox{.}(2022)]%
        {wang2022group}
\bibfield{author}{\bibinfo{person}{Dakuo Wang}, \bibinfo{person}{Haoyu Wang}, \bibinfo{person}{Mo Yu}, \bibinfo{person}{Zahra Ashktorab}, {and} \bibinfo{person}{Ming Tan}.} \bibinfo{year}{2022}\natexlab{}.
\newblock \showarticletitle{Group chat ecology in enterprise instant messaging: How employees collaborate through multi-user chat channels on slack}.
\newblock \bibinfo{journal}{\emph{Proceedings of the ACM on Human-Computer Interaction}} \bibinfo{volume}{6}, \bibinfo{number}{CSCW1} (\bibinfo{year}{2022}), \bibinfo{pages}{1--14}.
\newblock


\bibitem[Young et~al\mbox{.}(2006)]%
        {young2006don}
\bibfield{author}{\bibinfo{person}{Sandra~Florand Young}, \bibinfo{person}{Julia Hendry}, {and} \bibinfo{person}{Douglas Bicknese}.} \bibinfo{year}{2006}\natexlab{}.
\newblock \bibinfo{booktitle}{\emph{Don't Throw it Away!: Documenting and Preserving Organizational History}}.
\newblock \bibinfo{publisher}{Special Collections Department, The University Library and Jane Addams Hull~…}.
\newblock


\bibitem[Zhang and Cranshaw(2018)]%
        {zhang2018making}
\bibfield{author}{\bibinfo{person}{Amy~X Zhang} {and} \bibinfo{person}{Justin Cranshaw}.} \bibinfo{year}{2018}\natexlab{}.
\newblock \showarticletitle{Making sense of group chat through collaborative tagging and summarization}.
\newblock \bibinfo{journal}{\emph{Proceedings of the ACM on Human-Computer Interaction}} \bibinfo{volume}{2}, \bibinfo{number}{CSCW} (\bibinfo{year}{2018}), \bibinfo{pages}{1--27}.
\newblock


\bibitem[Zhang et~al\mbox{.}(2017)]%
        {zhang2017wikum}
\bibfield{author}{\bibinfo{person}{Amy~X Zhang}, \bibinfo{person}{Lea Verou}, {and} \bibinfo{person}{David Karger}.} \bibinfo{year}{2017}\natexlab{}.
\newblock \showarticletitle{Wikum: Bridging discussion forums and wikis using recursive summarization}. In \bibinfo{booktitle}{\emph{Proceedings of the 2017 ACM Conference on Computer Supported Cooperative Work and Social Computing}}. \bibinfo{pages}{2082--2096}.
\newblock


\bibitem[Zhang et~al\mbox{.}(2024)]%
        {zhang2024benchmarking}
\bibfield{author}{\bibinfo{person}{Tianyi Zhang}, \bibinfo{person}{Faisal Ladhak}, \bibinfo{person}{Esin Durmus}, \bibinfo{person}{Percy Liang}, \bibinfo{person}{Kathleen McKeown}, {and} \bibinfo{person}{Tatsunori~B Hashimoto}.} \bibinfo{year}{2024}\natexlab{}.
\newblock \showarticletitle{Benchmarking large language models for news summarization}.
\newblock \bibinfo{journal}{\emph{Transactions of the Association for Computational Linguistics}}  \bibinfo{volume}{12} (\bibinfo{year}{2024}), \bibinfo{pages}{39--57}.
\newblock


\end{thebibliography}

\end{document}